\documentstyle[12pt]{article}
\newcommand{\bee}{\begin{equation}}
\newcommand{\ee}{\end{equation}}
\newcommand{\tbeta}{\tilde\beta}
\newcommand{\bbeta}{\bar\beta}
\newcommand{\R}{\rm I\kern-.2emR}
\newcommand{\C}{\rm \kern.25em\vrule height1.4ex
 depth-.12ex width.06em\kern-.31em C}
\newcommand{\N}{{\rm I\kern-.16em N}}
\newcommand{\Z}{{\rm Z\kern-.35em Z}}
\begin{document}
\thispagestyle{empty}
\begin{flushright}
MPI-PhT/94-44\\
June 1994
\end{flushright}
\bigskip\bigskip\begin{center}
{\bf \Huge{Nonuniformity of the $1/N$ Expansion for
Two-Dimensional $O(N)$ Models}}
\end{center}  \vskip 1.0truecm
\centerline{\bf
Adrian Patrascioiu${}^*$ and Erhard Seiler}
\vskip5mm
\centerline{Max-Planck-Institut f\"{u}r
 Physik, Werner-Heisenberg-Institut}
\centerline{F\"ohringer Ring 6, 80805 Munich, Germany}
\vskip 2cm
\bigskip \nopagebreak \begin{abstract}
\noindent
We point out that the $1/N$ expansion, which is
widely invoked to infer properties of the $2D$ $O(N)$ models, is
nonuniform in the temperature, i.e. with decreasing temperature
the $1/N$ expansion truncated at a fixed order deviates more and
more from the true answer. This fact precludes the use of the expansion
to deduce low temperature properties such as asymptotic scaling for those
models. By contrast, in the $1D$ $O(N)$ chains, there are no signs of such
a nonuniformity.
\end{abstract}
\vskip 3cm
\bigskip \nopagebreak \begin{flushleft} \rule{2 in}{0.03cm}
\\ {\footnotesize \ ${}^*$ Permanent address: Physics Department, University
of Arizona, Tucson AZ 85721, \hglue 5mm U.S.A.}
\end{flushleft}

\newpage\setcounter{page}1

\section{Introduction}
The $1/N$ expansion \cite{Campo} is a very popular tool to analyze
$2D$ $O(N)$ models, especially in the critical dimension 2.
But already the first rigorous paper by A.Kupiainen
\cite {Kupi}, which established the fact that the expansion is a valid
asymptotic expansion in the thermodynamic limit at fixed
$\tbeta=\beta/N$, contained hints that the expansion is really only
good at high enough
temperature: the main result stated that the expansion is
valid provided $\tbeta$ is less than the critical value of the `spherical
model', i.e. the $N\to\infty$ limit of the model. Likewise the followup paper
by Fr\"ohlich et al \cite{FMR} proved Borel summability of the expansion
only for small $\tbeta$. M\"uller, Raddatz and R\"uhl \cite{MRR} computed
the $1/N$ corrections to the mass gap and the susceptibility
for large $\tbeta$; their
results are nonuniform in $\tbeta$ in the sense that the corrections
grow beyond all bounds as $\tbeta\to\infty$. This fact, which is
apparent from the equations in \cite{MRR} was, however, not noted explicitely
by the authors.
Finally, U.Wolff computed the correlation length  of the $O(N)$ model for
$N=2,3,4,8$. Comparing his results with the spherical model makes it
quite obvious that the approach to the $N\to\infty$ limit slows down with
increasing $\tbeta$ (see Tab.1).

In spite of all those facts, many authors (e.g.\cite{Campo,Flyv,BCR})
continue to cite the
$1/N$ expansion and the properties of the spherical model as evidence
for the supposed asymptotic scaling and absence of a critical point
of the $2D$ $O(N)$ model at finite $N$. In this note we report a few more
calculations that show explicitly that the $O(1/N)$ corrections for
a number of quantities grow linearly with $\tbeta$. All those
quantities are of long range character, such as correlation lengths,
susceptibilities or two-point functions at large distances. We did not
see such effects in short range quantities such as the energy (= two point
function at one lattice distance); this does not mean, however, that the
$1/N$ expansion for short range quantities is uniform in $\tbeta$.
Uniformity would require control of the deviation of the true result
from its truncated $1/N$ expansion, something that is not analyzed here.
We should, however, point out that our recent results on the problems of
perturbation theory (PT) \cite{superin} in non-Abelian models suggest
that in fact there is nonuniformity even in those short-distance quantitities.

The situation is different in $1D$: no indication of any non-uniformity
is found. The underlying reason is that in $1D$ the necessary computations
can be reduced to the analysis of simple 1 dimensional integrals, so
that no infrared effects, which in our mind are responsible for the
nonuniformity in $2D$, can arise.

\section{Notation and Definitions}

Consider the lattice $\Z^D$ and let $\Lambda$ be some given finite
subset of it. The non-linear $O(N)$ $\sigma$-model with standard nearest
neighbor interaction (s.n.n.i.) at inverse
temperature $\beta=\tbeta N$ is defined by the generating function

\bee
  Z_\Lambda(g)=\int\prod_{\langle ij\rangle}
\exp(\beta s_i\cdot s_j)\prod_i
\biggl(\exp(g_i\cdot s_i)\delta(s_i^2-1)ds_i\biggr)
\ee
or equivalently the Gibbs measure

\bee
 {1\over Z_\Lambda(0)}\prod_{\langle ij\rangle}\exp(\beta s_i\cdot s_j)
\prod_i\biggl(\delta(s_i^2-1)ds_i\biggr)
\ee
Here $s$ is a $N$-vector, the integration is over the spins inside $\Lambda$
with some boundary conditions.

It is well known (see for instance \cite{Kupi}) how one can develop
an asymptotic expansion in powers of $1/N$ by introducing a `Langrange
multiplier field' $a$ dual to the $\delta$-function constraint and
a standard saddle point expansion. In terms of the Lagrange multiplier
field one obtains a dual representation of the model; the generating
function is now given by
\bee
Z(g)=\int \prod_i da_i\exp(-N\hat S) \exp\biggl({1\over 2\tbeta N}
\bigl (g,(-\Delta+m^2-{2ia\over\sqrt{N}})^{-1} g\bigr)\biggr)
\ee
where the dual action is given by

\bee
\hat S={1\over 2}{\rm tr}\ln(-\Delta+m^2-
{2ia\over\sqrt{N}})-i\tbeta\sum_i{a_i\over\sqrt{N}}
\ee
The mass parameter $m$ appearing in this equation is fixed by the
gap equation

\bee
\tbeta=(-\Delta+m^2)^{-1}_{ii}
\ee
Kupiainen \cite{Kupi} showed, using the technique of cluster expansions,
that the resulting asymptotic expansion is uniform in the volume $\Lambda$
and can be used to obtain a valid asymptotic $1/N$ expansion in the infinite
volume, provided $\tbeta<\tbeta_{sph}$, where $\tbeta_{sph}$
is the critical point of the spherical model. It is well known how the
$1/N$ expansion can be cast in the form of Feynman graphs involving two
types of propagators $B$ and $C$ given by

\bee
C_{ij}=(-\Delta+m^2)^{-1}_{ij}
\ee
\bee
B_{ij}^{-1}={1\over 2}\bigl(C_{ij}^2\bigr)
\ee

Those Feynman graphs can be easily evaluated numerically on finite lattices.
In this paper we are considering the $1/N$ corrections to the correlation
length $\xi$, the susceptibility $\chi$ and the following ratio of two-point
functions

\bee
A\equiv {G(\xi_\infty/2)\over G(\xi_\infty/4)}
\ee
where $\xi_\infty$ is the correlation length of the spherical model (up
to a correction that is negligible for $\tbeta>.5$, $\xi_\infty=1/m$ where
$m$ is the solution of the gap equation) and

\bee
G(x)\equiv\langle s(0)\cdot s(x)\rangle
\ee
The quantities $\xi$, $\chi$, $A$ as well as the energy $E$
are expanded as follows:

\bee
\xi=\xi_{\infty}\bigl(1+{1\over N}\xi_1+O({1\over N^2})\bigr)
\ee
\bee
\chi=\chi_{\infty}\bigl(1+{1\over N}\chi_1+O({1\over N^2})\bigr);
\ee
\bee
{G({1\over 2}\xi_\infty)\over G({1\over 4}\xi_\infty)}=
A_\infty\bigl(1+{1\over N}a_1+O({1\over N^2})\bigr)
\ee

\bee
E\equiv\langle s(0)\cdot s(1)\rangle=E_\infty(1+{1\over N}e_1+O({1\over N^2})
\ee
In the following we will study the behavior of the coefficients $\xi_1$,
$\chi_1$ and $a_1$ for large $\tbeta$.
\section{What is the problem?}

Let us consider a certain physical quantity $A(N,\tbeta)$. According
to the preceding discussion it will have an asymptotic expansion
of the form

\bee
A(N,\tbeta)\sim\sum_{k=0}^\infty {a_k(\tbeta)\over N^k}
\ee
The meaning of the symbol $\sim$ is that the series on the right hand
side, if truncated at the term of order $N^{-k}$, will approximate the
left hand side by up to a remainder term which is $o(N^-k)$, and this is true
for any $k\geq0$. That means

\bee
r_k(N,\tbeta)\equiv A(N,\tbeta)-\sum_{k=0}^k{a_k(\tbeta)\over N^k}
\ee
satisfies
\bee
\lim_{N\to\infty}|r_k(N,\tbeta)|N^k=0
\ee

Uniform asymtoticity means the stronger property

\bee
|r_k(N,\tbeta)|N^k\leq c_k(N)
\ee
where $c_k(N)$ is independent of $\tbeta$ and satisfies
\bee
\lim_{N\to\infty}c_k=0
\ee

It is easy to see that uniform asymtopticity does not hold
if any one of the expansion coefficients $a_k(\tbeta)$ is
unbounded for large $\tbeta$, but it should be stressed that this is
not a necessary condition. There are simple examples of functions
where every one of the coefficients $c_k(\tbeta)$ is bounded
and yet the expansion is not uniformly asymptotic. Consider
for instance the function
\bee
f(N\tbeta)=\tbeta\exp(-N/\tbeta)
\ee
Its asymptotic expansion in powers of $1/N$ has all coefficients
$a_k(\tbeta)=0$ and the remainder term is always equal to $f$, which
is unbounded.

\section{One Dimension}

In $1D$ the situation is particularly simple because the thermodynamic
limit of the expectation values of local quantities can be obtained
by computing in a finite box with free boundary conditions.
For nearest neighbor quantities like the energy $E$ one only has to
evaluate a $1D$ integral; long range quantities such as the correlation
length $\xi$ or the susceptibility $\chi$ can be expressed in terms
of $E$. It turns out that for this reason no nonuniformity in any of
those quantities is found.

The energy of a $1D$ $O(N)$ chain can be obtained as follows:

\bee
E\equiv \langle s(0)\cdot s(1)\rangle={d\over d\beta}\ln Z
\ee
where $Z$ is the partition function for one link:

\bee
Z=\int_{-1}^1dte^{\beta t}(1-t^2)^{N-3\over 2}
\ee
Instead of using the general scheme outlined in the previous section, it is
convenient to obtain the $1/N$ expansion directly from this expression.
The two point function in the thermodynamic limit is simply given by

\bee
\langle s(0)\cdot s(n)\rangle=E^n
\ee
and hence the correlation length is

\bee
\xi=-\ln E
\ee
and the susceptibility

\bee
\chi={1+E\over 1-E}
\ee

Introducing the rescaled inverse temperature

\bee
\bbeta={\beta\over N-3}
\ee
the partition function takes the form

\bee
Z=\int_{-1}^1dte^{(N-3)f(t)}
\ee
with

\bee
f(t)=\bbeta t+{1\over 2}\ln(1-t^2)
\ee
It is now straightforward to produce a saddle point expansion in powers
of $1/(N-3)$: the saddle point value $t_o$ is given by

\bee
t_o(\bbeta)=\sqrt{1+{1\over 4\bbeta^2}}-{1\over 2\bbeta}
\ee
and the logarithm of the partition function is expanded as

\bee
\ln Z=(N-3)f_{\bbeta(t_o)}+{1\over 2}\ln{2\pi\over (N-3)(-f''_{\bbeta(t_o)})}
+O({1\over (N-3)^2})
\ee
Since $t_o$ is a function that increases monotonically from 0 to 1
as $\bbeta$ increases from 0 to $\infty$, we may use $t_o$ instead
of $\bbeta$ as an independent variable. Using

\bee
{d\over d\bbeta}={(1-t_o^2)^2\over 1+t_o^2}{d\over dt_o}
\ee
the expansion for the energy becomes
\bee
E=t_o-{t_o\over N-3}{(1-t_o^2)(3+2t_o^2)\over 1+t_o^2}
+O({1\over (N-3)^2})
\ee
It is already visible that the ration of the $1/(N-3)$ correction
to the leading term remains bounded as $\bbeta\to\infty$, i.e. $t_o\to 1$;
in fact it is $O(1/\bbeta)$.
But we want to expand in powers of $1/N$, not $1/(N-3)$, hence we have to
re-expand the leading term, using

\bee
\tbeta={\bbeta\over 1-3/N}
\ee
Again it is very easy to see that we just obtain another correction
of order $1/\tbeta$ to the $1/N$ contribution $e_1$ (see eq.(13)).
It is also easy to see that the result is in agreement with the one
found by Hasenfratz \cite{Has} for the infinite chain.

Using this information and  the formulae above
it is now trivial to conclude that

\bee
\xi_1={e_1\over\ln E}
\ee

\bee
\chi_1=-{2e_1E^2\over (1+E)(1-E)}
\ee

\bee
a_1=-{1\over 2}e_1\ln E
\ee
are all $O(1)$ for $\tbeta\to\infty$ as claimed.
\section{Two Dimensions}

We computed the Feynman graphs for the $1/N$ correction to the two point
function numerically on lattices satisfying $L/\xi_\infty=7$. It is
observed that this is sufficient to reach the thermodynamic limit with
a precision of 1 to 2\%.

The expansion for the two point function has the form

\bee
G(i)=G_\infty(i)+
{1\over N}\sum_{j,l\in\Lambda}{1\over L^4}
G_\infty(i-j)\Sigma(j-l) G_\infty(l)
\ee
where $G_\infty(i)=C(i)/\tbeta$ and
the `self-energy' $\Sigma$ is given by

\bee
\Sigma(i)=-B(i)G_\infty (i)+{1\over 2}\sum_{j,l,k\in\Lambda}{1\over L^6}
B(i-j)B(l-k)G_\infty (l-k)G_\infty (l-j)G_\infty (k-j)
\ee
and analogously in momentum space

\bee
\hat G(p)=\hat G_\infty(p)^2\hat\Sigma(p)
\ee
where $\hat\Sigma$ is given by

\bee
\hat\Sigma(p)=-\sum_k{1\over L^2}\hat G_\infty(k)\hat B(p-k)
+{1\over 2}\hat B(0)\sum_{k,q}{1\over L^4}\hat G_\infty(k)^2
\hat G_\infty(k+q)\hat B(q)
\ee
the momenta $p,k,q$ range over $2\pi i/L$, $i=1,2,...L$.
{}From the momentum space two-point functions $\hat G(0)$ and $\hat G(2\pi/L)$
we obtain the susceptibility and also the (effective) correlation length using
the formula

\bee
\xi= {1\over 2 \sin{\pi\over L}}
\sqrt{{\hat G(0)\over \hat G({2\pi\over L})}-1}
\ee
which is expanded in powers of $1/N$ using the expansion for the
two-point functions.

The results are given in Table 2. They show clearly that the $1/N$ corrections
for all three quantities grow linearly with $\tbeta$, respectively $\ln m$,
in agreement with the findings of \cite{MRR}.

\section{Discussion}

We have found by explicit calculation that the $1/N$ expansion
is not uniformly asymptotic in $\tbeta$ for a variety of
long range quantities in the $2D$ $O(N)$ model. The nonuniformity appears
in the most blatant form: the correction grows linearly with $\tbeta$.
For short distance quantities like the energy we could not find this kind of
phenomenon up to $O(1/N^3)$. This does not mean, however, that the $1/N$
expansion in uniform for those quantities; to show uniformity one would have to
control the remainder terms.

This situation is in striking contrast with the one for the $1D$ chains:
since all compututations reduce to one-dimensional integrals, there
is no nonuniformity occurring.

Some authors (for instance \cite{Flyv}) try to improve the quality
of the $1/N$ expansion by introducing a certain `mass renormalization'.
Since we are working on a lattice, there is first of all no need to do so
since there are no divergencies to be absorbed. Furthermore, the procedure
corresponds to a a resummation of certain terms in the $1/N$ expansion
and therefore is no longer a systematic asymptotic expansion in any parameter.
While apparently improving the agreement with the Monte Carlo data in the
$\tbeta$ region studied, the mathematical status of the procedure remains
unclear and it can be expected to reveal the same problems as the systematic
expansion at larger $\tbeta$.

In \cite{superin} we found that PT for the energy yields results that
depend on the boundary conditions to order $1/\tbeta^2$; this dependence
survives the limit $N\to\infty$. In our mind this
is a strong indication that the $1/N$ expansion is nonuniform in $\tbeta$ even
for short distance quantities, since the spherical limit of the model
does not show this phenomenon.

But the main conclusion is that that the nonuniformity for the mass gap
and similar quantities implies that it also holds for the Callan-Symanzik
$\beta$-function. This means that the $1/N$ expansion cannot be used to
draw any conclusions about the phase structure of the model for finite
$N$, in particular the fact that the spherical model does not have a soft
phase in $2D$ is in no way in conflict with the arguments advanced by us
\cite{Patr,PS} that strongly suggest the existence of such a phase for any
finite $N$.


\newpage

\noindent
{\bf Table 1}: {\it Comparison of Monte Carlo results with the spherical
limit};
Monte Carlo data taken from \cite{Gupta}($O(2)$), \cite{Apo} ($O(3)$),
\cite{Sokal} ($O(4)$) and\cite{Wolff} ($O(4)$ and $O(8)$). The numbers
represent $\xi/\xi_\infty$.

  \medskip
  \vbox{\offinterlineskip\halign{
  \strut\vrule#&\quad $#$\quad&\vrule\hskip1pt\vrule#&&\quad $#$\hskip5pt
  &\vrule#\cr
   \noalign{\hrule}
   &\tbeta&&.500&&.5667&&.575&&.600&&.650&&.725&\cr
   \noalign{\hrule\vskip1pt\hrule}
   &N=8&&1.307&&  &&1.494&&  &&1.711&&1.983&\cr
   \noalign{\hrule}
   &N=4&&1.866&&  &&2.857&&3.301&&4.425&&  &\cr
   \noalign{\hrule}
   &N=3&&2.644&&5.479&&8.348&&15.629&&  &&  &\cr
   \noalign{\hrule}
   &N=2&&9.691&&\infty&&\infty&&\infty&&\infty&&\infty&\cr
   \noalign{\hrule}}}

\vskip2cm

\noindent
{\bf Table 2}: {\it The $1/N$ corrections to various long range quantities}.
For the definition of the quantities given see eq. (10),(11),(12).

  \medskip
  \vbox{\offinterlineskip\halign{
  \strut\vrule#&\quad $#$\quad&\vrule\hskip1pt\vrule#&&\quad $#$\hskip5pt
  &\vrule#\cr
   \noalign{\hrule}
   &\tbeta&&.49349&&.60600&&.67105&&.71704&&.78173&&.82758&\cr
   \noalign{\hrule\vskip1pt\hrule}
   &L&&28&&56&&84&&112&&168&&224&\cr
   \noalign{\hrule}
   &\xi_\infty&&4&&8&&12&&16&&24&&32&\cr
   \noalign{\hrule}
   &\xi_1&&1.764&&2.927&&3.640&&4.152&&4.879&&5.397&\cr
   \noalign{\hrule}
   &\chi_1&&2.860&&4.932&&6.229&&7.171&&8.518&&9.487&\cr
   \noalign{\hrule}
   &a_1&&-.486&&-.853&&-1.058&&-1.205&&-1.416&&-1.567&\cr
   \noalign{\hrule}}}


\newpage
\begin{thebibliography}{1234567}
\newcommand{\bibi}[1]{\bibitem{#1}}
\newcommand{\authors}[1]{#1, }
\newcommand{\journal}[1]{{\sl #1}}
\newcommand{\volume}[1]{{\bf #1}}
\newcommand{\myyear}[1]{(#1)}
\newcommand{\page}[1]{#1}
\newcommand{\mytitle}[1]{{\it}}
\newcommand{\keywords}[1]{}

\bibi{Campo}
\authors{M.Campostrini and P.Rossi}
\journal{Riv.Nuovo Cim.}
\volume{16} \myyear{1993} \page{1}

\bibi{Kupi}
\authors{A.Kupiainen}
\journal{Commun.Math.Phys.}
\volume{73} \myyear{1980} \page{273}

\bibi{Has}
\authors{P.Hasenfratz}
\journal{Phys.Lett.}
\volume{141B} \myyear{1984} \page{385}

\bibi{FMR}
\authors{J.Fr\"ohlich, A.Mardin and V.Rivasseau}
\journal{Commun.Math.Phys.}
\volume{86} \myyear{1982} \page{87}

\bibi{MRR}
\authors{V.F.M\"uller, T.Raddatz and W.R\"uhl}
\journal{Nucl.Phys.}
\volume{B251[FS13]} \myyear{1985} \page{212}


\bibi{Flyv}
\authors{H.Flyvbjerg and S.Varsted}
\journal{Nucl.Phys.}
\volume{B344} \myyear{1990} \page{646};
\authors{H.Flyvbjerg and F.Larsen}
\journal{Phys.Lett.}
\volume{B266} \myyear{1991} \page{92};
\authors{H.Flyvbjerg and F.Larsen}
\journal{Phys.Lett.}
\volume{B266} \myyear{1991} \page{99};

\bibi{BCR}
\authors{P.Biscari, M.Campostrini and P.Rossi}
\journal{Phys.Lett.}
\volume{B242]} \myyear{1990} \page{225}

\bibi{superin}
\authors{A.Patrascioiu and E.Seiler}
{\it Superinstantons and the Reliability of Perturbation Theory in
Non-Abelian Models} preprint MPI-Ph/93-87 and AZPH-TH/93-33

\bibi{Patr}
\authors{A. Patrascioiu}
\mytitle{Existence of Algebraic Decay in non-Abelian Ferromagnets}
AZPH-TH/91-49

\bibi{PS}
\authors{A.Patrascioiu and E.Seiler}
\journal{Nucl.Phys.(Proc.Suppl.}
\volume{B30} \myyear{1993} \page{184}

\bibi{Gupta}
\authors{R.Gupta and C.F.Baillie}
\journal{Phys.Rev.}
\volume{B45} \myyear{1992} \page{2883}

\bibi{Apo}
\authors{J.Apostolakis, C.F.Baillie and G.F.Fox}
\journal{Phys.Rev.}
\volume{D43} \myyear{1991} \page{2687}

\bibi{Sokal}
\authors{R.G.Edwards, S.J.Ferreira, J.Goodman and A.D.Sokal}
\journal{Nucl.Phys.}
\volume{B380} \myyear{1992} \page{621}

\bibi{Wolff}
\authors{U.Wolff}
\journal{Phys.Lett.}
\volume{B248} \myyear{1990} \page{335}

\end{thebibliography}
\end{document}